\documentclass[a4paper,UKenglish,cleveref, autoref]{lipics-v2019}

\title{A compressed classical description of quantum states} 

\author{David Gosset}{IBM T.J. Watson Research Center, USA \and Department of Combinatorics and Optimization and Institute for Quantum Computing, University of Waterloo, Canada}{dgosset@uwaterloo.ca}{}{}

\author{John Smolin}{IBM T.J. Watson Research Center, USA}{smolin@us.ibm.com}{}{}

\authorrunning{D. Gosset and J. Smolin}

\Copyright{David Gosset and John Smolin}



\funding{We acknowledge support from the IBM Research Frontiers Institute.}

\acknowledgements{ DG thanks Scott Aaronson, Sergey Bravyi, Aram Harrow, Shelby Kimmel, Yi-Kai Liu, Ashley Montanaro, and Oded Regev for helpful discussions.
We thank Sergey Bravyi for his Clifford-manipulation code, and Robin Kothari for pointing out the relevance of the lower bound from  Ref.~\cite{gavinsky2007exponential}.}

\nolinenumbers 

\hideLIPIcs  

\EventEditors{John Q. Open and Joan R. Access}
\EventNoEds{2}
\EventLongTitle{42nd Conference on Very Important Topics (CVIT 2016)}
\EventShortTitle{CVIT 2016}
\EventAcronym{CVIT}
\EventYear{2016}
\EventDate{December 24--27, 2016}
\EventLocation{Little Whinging, United Kingdom}
\EventLogo{}
\SeriesVolume{42}
\ArticleNo{23}

\begin{document}

\maketitle

\begin{abstract}
We show how to approximately represent a quantum state using the square root of the usual amount of classical memory. The classical representation of an $n$-qubit state $\psi$ consists of its inner products with $O(\sqrt{2^n})$ stabilizer states. A quantum state initially specified by its $2^n$ entries in the computational basis can be compressed to this form in time $O(2^n \mathrm{poly}(n))$, and, subsequently, the compressed description can be used to additively approximate the expectation value of an arbitrary observable. Our compression scheme directly gives a new protocol for the \textit{vector in subspace problem} with randomized one-way communication complexity that matches  (up to polylogarithmic factors) the optimal upper bound, due to Raz. We obtain an exponential improvement over Raz's protocol in terms of computational efficiency.
\end{abstract}

\section{Compressing a quantum state}

A pure $n$-qubit state $\psi$ is fully specified by its $2^n$ amplitudes in the computational basis. Here we are interested in a compressed representation of size $\ll 2^n$  that can be used to approximately recover some of its features.

Aaronson described a compressed representation that can be used to approximate the expectation value of any observable from a known set \cite{aaronson2004limitations}. He showed that for any $n$-qubit state $\psi$ and any finite set $S$ of $n$-qubit observables, there is a compressed representation of $\psi$ of size $O(n\log(n)\log(|S|))$ which suffices to additively approximate expectation values of any observable in $S$. So for example even if $|S|=\exp(\mathrm{poly}(n))$, then one obtains a remarkable exponential reduction in the classical description size! \footnote{We note that a different compression method which achieves similar performance to Aaronson's   (in terms of the parameters described above) can be inferred from Ref. \cite{brandao2017exponential}. In this case the compressed representation is a classical description of the state produced at the output of Algorithm 5.} There are two limitations of Aaronson's scheme that we address here. The first is its efficiency: the algorithm used to compress an $n$-qubit quantum state scales as $\Omega(c^n|S|)$ for some constant $c>2$, which can be impractical. A second drawback is that one must first fix the set of observables $S$ and the compressed representation of $\psi$ then depends on this set. In this sense the compressed representation cannot be viewed as an unbiased description of $\psi$. 

Below we present a compressed representation which can be computed quickly, that is, in $O(2^n\mathrm{poly(n)})$ time, given the $2^n$ amplitudes which fully describe a quantum state $\psi$. This renders our method practical.  Moreover, we do not fix a set of observables a priori--it is possible to approximate the expectation value of an \textit{arbitrary} observable with high probability.

We note that Montanaro has recently ruled out the possibility of a more powerful kind of compressed representation of $\psi$ that would allow one to sample from the probabilty distribution obtained by measuring an observable  \cite{montanaro2016quantum}.

A compression scheme of the type we consider has two components. First, there is a classical \textit{compression algorithm} which takes an $n$-qubit state $\psi$ specified by its amplitudes in the computational basis, along with two parameters $\epsilon,p \in (0,1)$, and computes the compressed representation which we denote $D(\psi,\epsilon,p)$. The compression algorithm may be probabilistic in which case $D(\psi,\epsilon,p)$ is a random variable. Secondly, there is a classical \textit{expectation value algorithm} which takes as input $D(\psi,\epsilon,p)$ along with an $n$-qubit Hermitian operator $M$ satisfying $\|M\|\leq 1$, and outputs an estimate $E$ such that 
\begin{equation}
|E-\langle \psi|M|\psi\rangle|\leq \epsilon
\label{eq:E}
\end{equation}
with probability at least $1-p$, over the randomness used in both the compression and expectation value algorithms.  The \textit{classical description size} is the number of bits $|D(\psi,\epsilon,p)|$. We want this to be as small as possible. As we now explain, the best we can hope for is a classical description size which scales mildly exponentially with $n$.

Indeed, a limitation follows from related work concerning the communication complexity of the \textit{vector in subspace problem} \cite{kremer1995quantum, raz1999exponential, regev2011quantum}. Suppose Alice can send Bob information over a classical channel and that they may share a random bit string. Alice is given a (classical description) of an $n$-qubit quantum state $\psi$ and Bob is given an $n$-qubit projector $\Pi$. How much classical information does Alice need to send Bob so that he can compute the expectation value $\langle \psi|\Pi|\psi\rangle$? Suppose that we are promised that either $\langle \psi|\Pi|\psi\rangle\leq 1/3$ or $\langle \psi|\Pi|\psi\rangle\geq 2/3$, so that Bob's goal is to compute just this one bit of information; we allow him to err with probability at most $\frac{1}{4}$, say. The number of bits that Alice needs to send for them to jointly succeed at this task is called the one-way (randomized) classical communication complexity of the vector in subspace problem. 

Raz showed that Alice need only send $O(\sqrt{2^n})$ bits for them to succeed \cite{raz1999exponential}. He proposed the following simple protocol which achieves this bound. Using the shared random bit string Alice computes a set of $W=2^{2^{n/2}}$ Haar random $n$-qubit states $\phi_1,\phi_2,\ldots,\phi_{W}$ and then computes the inner products $\langle \phi_1|\psi\rangle,\ldots, \langle \phi_W|\psi\rangle$. She identifies the state $\phi_j$ such that $|\langle \phi_j|\psi\rangle|$ is maximal and sends the index $j$ to Bob, encoded as a bit string of length $2^{n/2}$. Using the index $j$ along with the shared random string Bob can compute the state $\phi_j$. He then computes $\langle \phi_j |\Pi|\phi_j\rangle$ and outputs 1 if it is larger than a certain threshold value and zero otherwise. A detailed analysis of this protocol is provided in Ref. \cite{montanaro2016quantum}.  Note that it is not a practical method of compressing a quantum state as its runtime is doubly exponential in $n$.   

Raz's protocol is optimal in terms of the number of bits communicated. Indeed, a matching lower bound of $\Omega(\sqrt{2^n})$  for the one-way classical communication complexity of the vector in subspace problem can be inferred from the work of Gavinsky et al. \cite{gavinsky2007exponential} (the following argument was suggested to us by R. Kothari). In that work the authors describe a communication task called the $1/4$-Partial Matching problem. Let $N=2^n$. In this problem Alice is given an $N$-bit string $x\in \{0,1\}^{N}$, and Bob is given an $N/4$-bit string $w\in \{0,1\}^{N/4}$ as well as a ``partial matching" which is a set of $N/4$ disjoint pairs
\[
\{(i_1,j_1),(i_2,j_2), \ldots, (i_{N/4}, j_{N/4})\}
\]
such that $1\leq i_k, j_k\leq N$ for all $k=1,2,\ldots, N/4$. We are promised that there exists a bit $b\in \{0,1\}$ such that 
\begin{equation}
w_k\oplus x_{i_k}\oplus x_{j_k}=b \qquad \quad 1\leq k\leq N/4.
\label{eq:match}
\end{equation}
The goal is to compute the bit $b$. Gavinsky et al. establish that the classical one-way communication complexity of the $1/4$-Partial Matching problem is $\Theta(\sqrt{N})$, and also provide a one-way quantum communication protocol using only $O(\log(N))$ qubits. As we now show, the latter protocol can be recast as a protocol for a specific class of instances of the vector in subspace problem with $n$ qubits. Let us write
\[
\mathcal{V}=\mathrm{span}\left\{|i_k\rangle, |j_k\rangle: 1\leq j,k\leq N/4\right\}
\]
for the $N/2$-dimensional subspace spanned by basis vectors contained in Bob's matching, and let $\mathcal{V}^{\perp}$ be the orthogonal subspace, which is also $N/2$-dimensional. Finally, let $P$ be a projector onto a subspace of $\mathcal{V}^{\perp}$ of dimension $N/4$ which is spanned by $N/4$ basis vectors (these can be any $N/4$ basis vectors which are not contained in Bob's matching).  Then consider the $n$-qubit state $\psi$ and projector $\Pi$ defined as follows.
\[
|\psi\rangle=\frac{1}{\sqrt{N}}\sum_{i=1}^{N} (-1)^{x_i}|i\rangle
\]
\[
\Pi=P+\sum_{k=1}^{N/4} \frac{1}{2}\left(|i_k\rangle-(-1)^{w_k} |j_k\rangle\right)\left(\langle i_k|-(-1)^{w_k}\langle j_k|\right)
\]
Using Eq.~\eqref{eq:match} and the definition of $P$ we see that $\langle \psi|\Pi|\psi\rangle=\frac{1}{4}(1+2b)$ for $b\in \{0,1\}$. Thus when $b=1$ we have  $\langle \psi|\Pi|\psi\rangle\geq 2/3$ and when $b=0$ we have $\langle \psi|\Pi|\psi\rangle\leq 1/3$.  This provides the desired reduction from the $1/4$-Partial Matching problem with $N=2^n$ to the vector in subspace problem with $n$ qubits and Ref. \cite{gavinsky2007exponential} then implies a lower bound $\Omega(\sqrt{2^n})$ on its one-way classical communication complexity. 

This lower bound directly provides a limitation on compression schemes of the type described above. Indeed, any such compression scheme can be trivially converted into a protocol for solving the vector in subspace problem, with one-way communication complexity $|D(\psi,\epsilon,p)|$ for $\epsilon,p=\Theta(1)$. \footnote{Given $\psi$, Alice computes $D(\psi, \epsilon=1/100 ,p=1/4)$ and sends it to Bob, who then uses it to compute an estimate $E$ such that $|E-\langle\psi|\Pi|\psi\rangle|\leq 1/100$ with probability at least $\frac{1}{4}$. Bob outputs $1$ if $E\geq 1/2$ and $0$ otherwise.} Thus any compression scheme must have classical description size scaling as $\Omega(\sqrt{2^n})$, so we can only hope to match Raz's upper bound in our slightly more general setting. Finally, although here we will not go beyond the one-way setting, we note that Ref.\cite{regev2011quantum} establishes an exponential lower bound on the total classical communication used to solve the vector in subspace problem allowing for multiple rounds of communication.

In the remainder we present our compression scheme for quantum states. We assume basic knowledge of quantum computation and write $\mathcal{C}_n$ for the $n$-qubit Clifford group (see, e.g., \cite{gottesman1998heisenberg}). Recall that an $n$-qubit unitary $C$ is an element of $\mathcal{C}_n$ if and only if it can be expressed as a quantum circuit composed of single-qubit Hadamard gates, single-qubit phase gates $S=\mathrm{diag}(1,i)$, and two-qubit CNOT gates. In the following we shall use the fact that (A) the group $\mathcal{C}_n$ of Clifford unitaries forms a unitary $2$-design \cite{dankert2009exact}, and  (B) every Clifford unitary has a short classical description as a quantum circuit consisting of $O(n^2)$ elementary gates.

Let $\psi$ be an $n$-qubit state. A \textit{stabilizer sketch} of $\psi$ of size $2^k$ is a classical description of a Clifford unitary $C\in \mathcal{C}_n$ along with the vector of $2^k$ amplitudes
\begin{equation}
\quad \quad \quad \langle 0^{n-k} z|C|\psi\rangle \qquad \qquad z\in \{0,1\}^k.
\label{eq:amps}
\end{equation}
A random stabilizer sketch of size $2^k$ is obtained by choosing $C\in \mathcal{C}_n$ uniformly at random. If $\psi$ is stored in computer memory as a list of $2^n$ amplitudes in the computational basis, then a random stabilizer sketch of $\psi$ can be computed as follows. The random Clifford $C$ can be drawn \cite{koenig2014efficiently} and expressed as a circuit consisting of $O(n^2)$ one- and two-qubit Clifford gates \cite{aaronson2004improved}, using a total runtime of $O(n^3)$. This circuit can be applied to $\psi$ to obtain $C|\psi\rangle$ using $O(2^n n^2)$ elementary arithmetic operations. We retain only $2^k$ computational basis amplitudes of this state along with the classical description of $C$.

We now show that a handful of random stabilizer sketches of $\psi$ can serve as a compressed representation of $\psi$. In particular, given parameters $\epsilon,p>0$, our compressed representation $D(\psi,\epsilon,p)$ consists of $O(\log(p^{-1}))$ independent random stabilizer sketches of  $\psi$, each of size 
\begin{equation}
2^k=\tilde{O}(\sqrt{2^n}\epsilon^{-1})
\label{eq:kval}
\end{equation}
 Here and below we use the $\tilde{O}$ notation to hide $\mathrm{poly}(n,\log(\epsilon^{-1}))$ factors, and it is sufficient that the amplitudes Eq.\eqref{eq:amps} provided in the stabilizer sketches are specified to $\tilde{O}(1)$ bits of precision. We shall also assume $\epsilon\geq 2^{-n/2}$, since otherwise the scaling from Eq.~\eqref{eq:kval} is no better than the trivial $O(2^n)$. Our scheme therefore achieves a classical description size:
\begin{equation}
\left|D(\psi,\epsilon,p)\right|=\tilde{O}\big(\sqrt{2^n}\epsilon^{-1} \log(p^{-1})\big),
\label{eq:dval}
\end{equation}
matching Raz and the lower bound from Ref.~\cite{gavinsky2007exponential} up to the factors hidden in the $\tilde{O}$. Moreover, the compression algorithm is to simply compute these random stabilizer sketches which takes time $\tilde{O}(2^n)$ as described above.

It remains to exhibit the expectation value algorithm, which takes as input such a compressed representation $D(\psi,\epsilon,p)$ along with an $n$-qubit observable $M$ and computes an approximation $E$ satisfying Eq.~\eqref{eq:E} with probability at least $1-p$.  The following lemma describes a slightly more general algorithm which has an improved performance if the observable $M$ has low rank; the claimed expectation value algorithm achieving Eqs. (\ref{eq:kval},\ref{eq:dval}) is the unrestricted case $r=2^n$ (note that Eq.~\eqref{eq:kchoice} matches Eq.~\eqref{eq:kval} whenever $\epsilon \geq 2^{-n/2}$).

\begin{lemma}
There is a deterministic classical algorithm which takes as input positive integers $n,r$ with $1\leq r\leq 2^n$ and real numbers $\epsilon,p\in (0,1)$, along with
\begin{itemize}
\item{An $n$-qubit Hermitian operator $M$ satisfying $\|M\|\leq 1$ and $\mathrm{rank}(M)\leq r$.}
\item{ $O(\log(p^{-1}))$ independent random stabilizer sketches of an $n$-qubit state $\psi$, each of size $2^k$, where $k$ is the largest integer satisfying 
\begin{equation}
2^k\leq 24\cdot \max\left\{\epsilon^{-2},\sqrt{r}\epsilon^{-1}\right\}.
\label{eq:kchoice}
\end{equation}
}
\end{itemize}
The algorithm outputs an estimate $E$ such that, with probability at least $1-p$
\[
|E-\langle \psi|M|\psi\rangle|\leq \epsilon.
\]
The runtime of the algorithm is $2^{O(n)}\log(p^{-1})$.
\label{lem:expectationalgorithm}
\end{lemma}
This completes our description of the compression scheme for quantum states. We pause for a few remarks before giving the proof below. 

As noted above, taking $r=2^n$ in the Lemma, our compression scheme gives a protocol for the vector in subspace problem. In the opposite extreme case where $r=1$ we have $M=|\phi\rangle\langle \phi|$ for some $n$-qubit state $\phi$. Here we obtain a one-way communication complexity protocol for  approximating the inner product between two vectors $\psi$ (Alice's input) and $\phi$ (Bob's input) to some additive error $\epsilon$.  Choosing $\epsilon=O(1)$, the communication complexity of this protocol is then $\tilde{O}(1)$ which is known to be optimal \cite{kremer1999randomized}.

The compressed representation can be used to simultaneously approximate expectation values of multiple observables with low probability of failure. Suppose we choose $p=\frac{\delta}{K}$ for some $\delta>0$ and positive integer $K$. By a union bound, with probability at least $1-\delta$,  the compressed representation can be used to compute estimates of the expectation values for all observables in any set $S$ of size $K$, such that all are within the desired approximation error $\epsilon$. Crucially, the compressed representation does not depend on the set $S$, which highlights the difference between our setting and the one considered by Aaronson \cite{aaronson2004limitations}.

Finally, an interesting special case is when the observable $M$ is a stabilizer projector, i.e., 
\[
M=C^{\dagger} \left( |0\rangle\langle 0|^{\otimes m}\otimes I_{n-m}\right) C
\]
for some Clifford $C\in \mathcal{C}_n$ and integer $0\leq m\leq n$. For example, to approximate the expectation value of any $n$-qubit Pauli $P$ it suffices to approximate the expected value of the stabilizer projector $(I+P)/2$. As we explain in more detail below, if $M$ is a stabilizer projector then the algorithm from the Lemma has an improved runtime of $\tilde{O}(r \log(p^{-1}))$ and only uses $\tilde{O}(\sqrt{r}\log(p^{-1}))$ space. 

We now present the proof of the lemma. 
\begin{proof}
Note that since $k$ is the largest positive integer satisfying Eq.\eqref{eq:kchoice}, we have
\begin{equation}
2^k \geq 12\cdot \max\left\{\epsilon^{-2},\sqrt{r}\epsilon^{-1}\right\}.
\label{eq:kchoice2}
\end{equation}

Suppose first that we are given just $2$ independent random stabilizer sketches of $\psi$ of size $2^k$. The two sketches comes with $n$-qubit Cliffords $C,D\in \mathcal{C}_n$; we define the associated stabilizer code projectors
\begin{equation}
P=C^{\dagger} \left(|0\rangle\langle0|_{n-k}\otimes I_k \right)C\qquad Q=D^{\dagger} \left(|0\rangle\langle0|_{n-k}\otimes I_k \right)D.
\label{eq:nk}
\end{equation}
Since $C,D$ are uniformly random, $P$ and $Q$ are uniformly random $n$-qubit stabilizer projectors of rank $2^k$. Therefore
\begin{equation}
\mathbb{E}[P]=\mathbb{E}[Q]=2^{k-n} I.
\label{eq:P}
\end{equation}
Using the fact that the $n$-qubit Clifford group is a unitary $2$-design \cite{dankert2009exact} one can also directly show the following fact (we provide a proof for completeness below). 
\begin{claim}
\begin{equation}
\mathbb{E}[P\otimes P]=\mathbb{E}[Q\otimes Q]= a\cdot I+b\cdot \mathrm{SWAP}
\label{eq:ptimesp}
\end{equation}
where
\begin{equation}
a=\frac{4^{k+n}-2^{k+n}}{4^{2n}-4^{n}}\leq 4^{k-n} \qquad \qquad b=\frac{4^{n}2^k-2^{n}4^k}{4^{2n}-4^{n}} \leq 2^k4^{-n}.
\label{eq:ab}
\end{equation}
\end{claim}
Here and below we write $\mathrm{SWAP}$ for the unitary which swaps two $n$-qubit registers, defined by its action on computational basis states: $\mathrm{SWAP}|x\otimes y\rangle=|y\otimes x\rangle$. 

From the two stabilizer sketches we may compute
\begin{align}
F& =4^{n-k}\mathrm{Re}\left(\langle \psi|P M Q|\psi\rangle\right)\label{eq:Fdef}\\
&=4^{n-k}\mathrm{Re}\bigg(\sum_{x,y\in \{0,1\}^k}\langle \psi|C^{\dagger}|0^{n-k}x\rangle\langle 0^{n-k}x|C M D^{\dagger}|0^{n-k}y\rangle\langle 0^{n-k}y|D|\psi\rangle\bigg).
\label{eq:expand}
\end{align}
Indeed, Eq. \eqref{eq:expand} expresses $F$ in terms of the amplitudes $\langle \psi|C^{\dagger}|0^{n-k}x\rangle, \langle 0^{n-k}y|D|\psi\rangle$ from the two given stabilizer sketches as well as matrix elements $\langle 0^{n-k}x|C M D^{\dagger}|0^{n-k}y\rangle$ which can be computed from the classical descriptions of Cliffords $C,D$ and the given observable $M$. Note that the computation of $F$ involves a summation of $2^{2k}$ terms and each term involves the computation of a matrix element $\langle 0^{n-k}x|C M D^{\dagger}|0^{n-k}y\rangle$, which takes time $2^{O(n)}$. Therefore the total time to compute $F$ given the two stabilizer sketches is $2^{O(n)}$. We note that in the special case where $M$ is a stabilizer projector the matrix element $\langle 0^{n-k}x|C M D^{\dagger}|0^{n-k}y\rangle$ is equal to the inner product between stabilizer states $\langle 0^{n-k}x|C$ and $M D^{\dagger}|0^{n-k}y\rangle$ and can be computed in time $O(n^3)$ (see e.g., Ref. \cite{bravyi2016improved}).  Thus in this case $F$ can be computed using time $\tilde{O}(2^{2k})=\tilde{O}(r)$ and space $\tilde{O}(\sqrt{r})$.

Using Eqs.~(\ref{eq:P},\ref{eq:Fdef}) we see that 
\begin{equation}
\mathbb{E}\left[F\right]=\langle \psi|M|\psi\rangle.
\label{eq:mean}
\end{equation}
We now bound the variance of $F$. First note that
\begin{equation}
4^{-2(n-k)}F^2\leq |\langle \psi|P MQ|\psi\rangle|^2.
\label{eq:F2}
\end{equation}
We use the identity 
\[
\mathrm{Tr}\left(\alpha \otimes \gamma \cdot \beta \otimes \delta  \cdot \mathrm{SWAP}\right)=\mathrm{Tr}\left(\alpha \beta\gamma \delta \right)
\]
which holds for square matrices $\alpha,\beta,\gamma,\delta$ (all of the same dimensions). Using this fact in Eq.~\eqref{eq:F2} we get
\begin{align}
4^{-2(n-k)}F^2&\leq \big|\mathrm{Tr}\big(\left(|\psi\rangle\langle \psi|\otimes M\right)\left( P\otimes Q \right) \mathrm{SWAP}\big)\big|^2\\
&=\mathrm{Tr}\big(\left(|\psi\rangle\langle \psi|\otimes M\otimes M \otimes |\psi\rangle \langle \psi|\right)\left( P\otimes Q\otimes P\otimes Q\right) \mathrm{SWAP}_{12}\mathrm{SWAP}_{34}\big).
\end{align}

Now taking the expectation value of both sides and using Eq.~\eqref{eq:ptimesp} we obtain
\[
4^{-2(n-k)}\mathbb{E}\left[F^2\right] \leq \mathrm{Tr}\big( \Gamma \cdot  \left(|\psi\rangle\langle \psi|\otimes M\otimes M \otimes |\psi\rangle \langle \psi|\right)\big)
\]
where
\[
 \Gamma=\big(a\cdot I+b \cdot \mathrm{SWAP}_{13}\big)\big(a\cdot I+b\cdot\mathrm{SWAP}_{24}\big)\mathrm{SWAP}_{12}\mathrm{SWAP}_{34}
\]
and $a,b$ are defined in Eq. \eqref{eq:ab}.  Evaluating the trace term by term we arrive at
\begin{align}
4^{-2(n-k)}\mathbb{E}\left[F^2\right]& \leq a^2 \left(\langle \psi|M|\psi\rangle\right)^2+2ab\langle\psi|M^2|\psi\rangle+b^2\mathrm{Tr}(M^2)\\
&\leq 4^{-2(n-k)} \left(\langle \psi|M|\psi\rangle\right)^2+2\cdot 2^{3k}4^{-2n}\langle\psi|M^2|\psi\rangle+4^{k-2n}\mathrm{Tr}(M^2)
\end{align}
where in the last line we used Eq. \eqref{eq:ab}.
Therefore
\begin{equation}
\mathrm{Var}(F)= \mathbb{E}\left[F^2\right]-\big(\mathbb{E}[F]\big)^2\leq \frac{2}{2^k}\langle\psi|M^2|\psi\rangle+\frac{1}{4^k} \mathrm{Tr}(M^2).
\label{eq:var}
\end{equation}
Using the bounds 
\[
\mathrm{Tr}(M^2)\leq \|M\|^2 \mathrm{rank}(M)\leq r,
\]
and $\langle \psi|M^2|\psi\rangle\leq \|M\|^2\leq 1 $ in Eq.\eqref{eq:var} gives
\begin{equation}
\mathrm{Var}(F)\leq \frac{2}{2^k}+\frac{r}{4^k}\leq \frac{\epsilon^2}{6}+\frac{\epsilon^2}{12^2}\leq \frac{\epsilon^2}{4},
\label{eq:var2}
\end{equation}
where we used Eq.~\eqref{eq:kchoice2}.

Putting together Eqs.~(\ref{eq:mean},\ref{eq:var2}) and applying Chebyshev's inequality we get
\[
\mathrm{Pr}\left[|F-\langle \psi|M|\psi\rangle|\geq \epsilon\right]\leq \frac{1}{4}.
\]

We have shown that just two stabilizer sketches of size $2^k$ suffice to compute an estimate $F$ which, with probability at least $3/4$, achieves the desired precision $\epsilon$. The success probability can be amplified by taking the median of multiple independent estimates \cite{jerrum1986random}.  That is, now using $2L$ random stabilizer sketches (of the same size $2^k$), we compute $L$ independent estimates $F_1,\ldots, F_L$ as above and compute the median
\[
E=\mathrm{Median}(F_1,F_2,\ldots, F_L).
\]
Note that if $|E-\langle \psi|M|\psi\rangle|\geq \epsilon$ then at least $\frac{1}{2}(L-1)$ of the estimates $F_j$ lie outside the interval $(\langle \psi|M|\psi\rangle-\epsilon, \langle \psi|M|\psi\rangle+\epsilon)$. Since these  events are independent and each occurs with probability at most $1/4$ we have
\begin{equation}
\mathrm{Pr}\left[ |E-\langle \psi|M|\psi\rangle|\geq \epsilon\right]\leq \sum_{\frac{1}{2}(L-1)\leq \ell\leq L}\binom{L}{\ell} \left(\frac{3}{4}\right)^{L-\ell} \left(\frac{1}{4}\right)^{\ell}.
\label{eq:median}
\end{equation}
By a Chernoff bound the right-hand side of Eq. \eqref{eq:median} can be made $\leq p$ by choosing $L=O(\log(p^{-1})$.

The algorithm which computes $E$ simply uses Eq.~\eqref{eq:expand} to compute the independent estimates $F_1,\ldots, F_L$ and then takes the median. The computation of each $F_j$ takes time $2^{O(n)}$ as discussed above.
\end{proof}

\begin{figure}[t]
\centering
	\begin{subfigure}{.32\textwidth}
		\includegraphics[width=\textwidth]{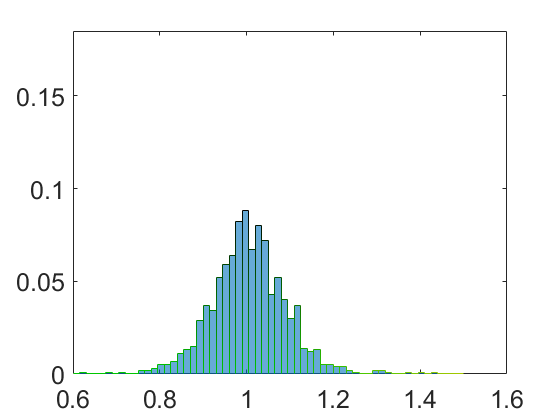}
	\end{subfigure}
	\begin{subfigure}{.32\textwidth}
		\includegraphics[width=\textwidth]{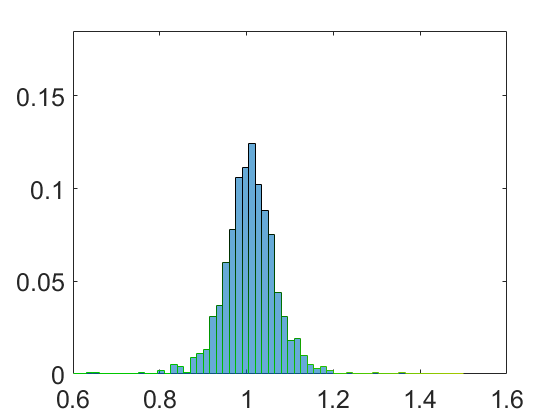}
	\end{subfigure}
	\begin{subfigure}{.32\textwidth}
		\includegraphics[width=\textwidth]{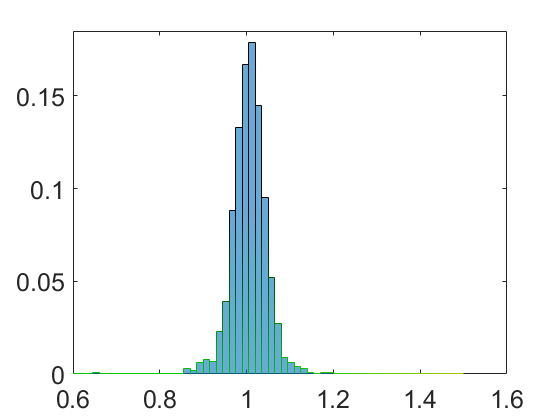}
	\end{subfigure}
	\caption{These histograms show the distribution of the random variable $F$ defined in Eq.~\eqref{eq:Fdef} in three  examples where $n=12$ and $k=7,8,9$ (left to right). Here $\psi$ was chosen to be a random state in the $12$-qubit symmetric subspace and $\Pi$ was chosen to be the projector onto this subspace, which has rank $r=13$. Thus $\mathbb{E}[F]=\langle \psi|\Pi|\psi\rangle=1$.  For each $k=7,8,9$ we computed $1000$ realizations of $F$. The upper bound on the standard deviation computed from Eq.~\eqref{eq:var} gives $ 0.1259, 0.0887, 0.0626$ respectively while the standard deviation of the samples was computed to be $0.0875,0.0637,0.0399$.  These figures were generated using Python libraries for Clifford manipulations which will soon be available in QISKit\cite{QISKit}.
\label{fig:sym}}
\end{figure}

This compression scheme may be useful as a memory-saving means of storing or transmitting the output of classical simulations of large quantum circuits.  Suppose the $n$-qubit output state $|\psi\rangle$ of such a simulation is compressed and sent to a client with at most $O(\sqrt{2^n})$ memory available on her local machine.  With this much memory the client cannot store the full state $\psi$ but she can store its compressed representation and use it to approximate the expectation value of any stabilizer projector (as discussed above). The amount of memory savings that could be achieved in practice is determined by the size $2^k$ of the stabilizer sketches which should be used for a given number of qubits $n$ and error $\epsilon$. Let us suppose for simplicity we are willing to accept a failure probability of $p=1/4$ so that only 2 stabilizer sketches are needed using the above scheme.  To obtain a more precise estimate of this $k$ than the one from  Eq.~\eqref{eq:kchoice},  we may solve for $\mathrm{Var}(F)\leq \epsilon^2/4$ using the upper bound on $\mathrm{Var}(F)$ from Eq.~\eqref{eq:var}. For example, for a very large state of $n=50$ qubits we may choose $k=35$ to obtain error $\epsilon=0.0039$. This corresponds to a memory savings of a factor of $2^{14}=16384$.

Fig. \ref{fig:sym} shows the distribution of the estimator $F$ from Eq.~\eqref{eq:Fdef} in three examples where $n=12$, $r=13$, and $k=7,8,9$. In all of these examples the standard deviation of the samples is within a factor of $2$ of the upper bound on the standard deviation computed from Eq.~\eqref{eq:var}, showing that this upper bound cannot be improved much further.

\paragraph{Proof of Claim 2}
The $n$-qubit Clifford group $\mathcal{C}_n$ is a unitary $2$-design \cite{dankert2009exact}. This means that if $C\in \mathcal{C}_n$ is uniformly random then for any $2n$-qubit operator $\rho$ we have
\[
\mathbb{E}\left[C^{\dagger} \otimes C^{\dagger}\rho C \otimes C\right]=\int_{Haar} dU \; U^{\dagger}\otimes U^{\dagger} \rho U\otimes U
\]
where the right-hand side is integrated over the Haar measure on the unitary group $U(2^n)$. Define projectors $\pi_{\pm}=(I\pm \mathrm{SWAP})/2$ onto the symmetric and antisymmetric subspaces of two $n$-qubit registers. Using Schur's lemma the right-hand side of the above expression can be further simplified to
\[
\int_{Haar} dU \; U^{\dagger}\otimes U^{\dagger} \rho U\otimes U= \pi_{+} \frac{\mathrm{Tr}(\rho\pi_{+})}{\mathrm{Tr}(\pi_+)}+\pi_{-} \frac{\mathrm{Tr}(\rho \pi_{-})}{\mathrm{Tr}(\pi_{-})}.
\]
Eq.~\eqref{eq:ptimesp} is obtained by applying this formula in the case $\rho=R\otimes R$, where $R=|0\rangle\langle 0|_{n-k}\otimes I_k$, and substituting
\[
\mathrm{Tr}(R\otimes R \cdot \pi_{\pm})=\frac{\mathrm{Tr}(R)^2\pm \mathrm{Tr}(R)}{2}=\frac{2^{2k}\pm 2^k}{2} \qquad \qquad \mathrm{Tr}(\pi_{\pm})=\frac{4^n\pm 2^n}{2}.
\]

\bibliography{sketch}

\end{document}